\documentclass[twocolumn,showpacs,amsmath,amssymb,prl,floatfix]{revtex4}

\usepackage{graphicx}
\usepackage{epsfig}
\usepackage{bm}

\usepackage{color}

\begin{document}

\title{Antiferromagnetic long-range order in dissipative Rydberg lattices}
\author{Michael Hoening$^1$, Wildan Abdussalam$^2$, Michael Fleischhauer$^1$, and Thomas Pohl$^2$ }
\affiliation{$^1$ Fachbereich Physik und Forschungszentrum OPTIMAS, Technische Universit\"at Kaiserslautern, D-67663 Kaiserslautern, Germany}
\affiliation{$^2$ Max Planck Institute for the Physics of Complex Systems, N\"othnitzer Stra\ss e 38, 01187 Dresden, Germany}

\begin{abstract}
We study the dynamics of dissipative spin lattices with power-law interactions, realized via few-level atoms driven by coherent laser-coupling and decoherence processes. Using Monte-Carlo simulations, we determine the phase diagram in the steady state and analyze the dynamics of its generation. As opposed to mean-field predictions and nearest-neighbour models there is no phase transition to long-range ordered phases for realistic interactions and resonant driving. However, for finite laser detunings, we demonstrate the emergence of crystalline order with a vanishing dissipative gap. Although the found steady states differ considerably from those of an equilibrium Ising magnet, the critical exponent of the revealed dissipative phase transition falls into the 2D Ising universality class. Two complementary schemes for an experimental implementation with cold Rydberg atoms are discussed. 
\end{abstract}


\maketitle
Experiments with cold atoms offer unique insights into many-body physics \cite{bdz08}. Measurement techniques have reached spatio-temporal resolution \cite{res_meas} that -- combined with tuneable zero-range interactions -- are opening the door for microscopic observations of dynamical phenomena. Laser-excitation of Rydberg states adds strong and long-range interactions to this powerful toolbox \cite{ryd_rev}, being key to the exploration of new ordered phases with ultracold atoms 
\cite{Lahaye2009}. One avenue is to prepare such phases as many-body ground states of a Hermitian Hamiltonian via slow changes of laser frequency and/or intensity \cite{pdl10,slm10,bsl11}. This requires sufficient time to remain adiabatic, which poses a challenging competition with the finite lifetime of the Rydberg states. Alternatively, this excited-state decay has been proposed as a natural means for dissipative state preparation \cite{lhc11}. The non-equilibrium physics of such driven open systems has recently attracted considerable interest \cite{Diehl2010,Mueller2012,Sieberer2013,Lee2013}, as their properties can differ dramatically from conventional equilibrium situations.

Previous work considered lattices of effective spins, represented by an atomic ground and strongly interacting Rydberg state coupled by external laser driving and spontaneous decay \cite{lhc11,Ates2012,hmp13}. The emergence of steady states with antiferromagnetic order was predicted on the basis of mean field theory assuming nearest-neighbour (NN) interactions \cite{lhc11}. 
It was later shown, however, that ordering in these systems is restricted to short length scales for all spatial lattice dimensions due to large single-site fluctuations associated with a simple two-level driving scheme \cite{hmp13}.  Simulations moreover showed that crystallization in 1D is precluded for other driving schemes \cite{hmp13}, for which long-range order was predicted using mean field theory \cite{Qian2012}. Hence, the possibility of long-range order in dissipative Rydberg lattices has thus far remained an open question.

%%%%%%%%%%%%%%%%%%%%%%%%%%%%%%%%%%%%%%%%%%%%%%%
\begin{figure}[t!]
\begin{center}
\resizebox{0.99\columnwidth}{!}{\includegraphics{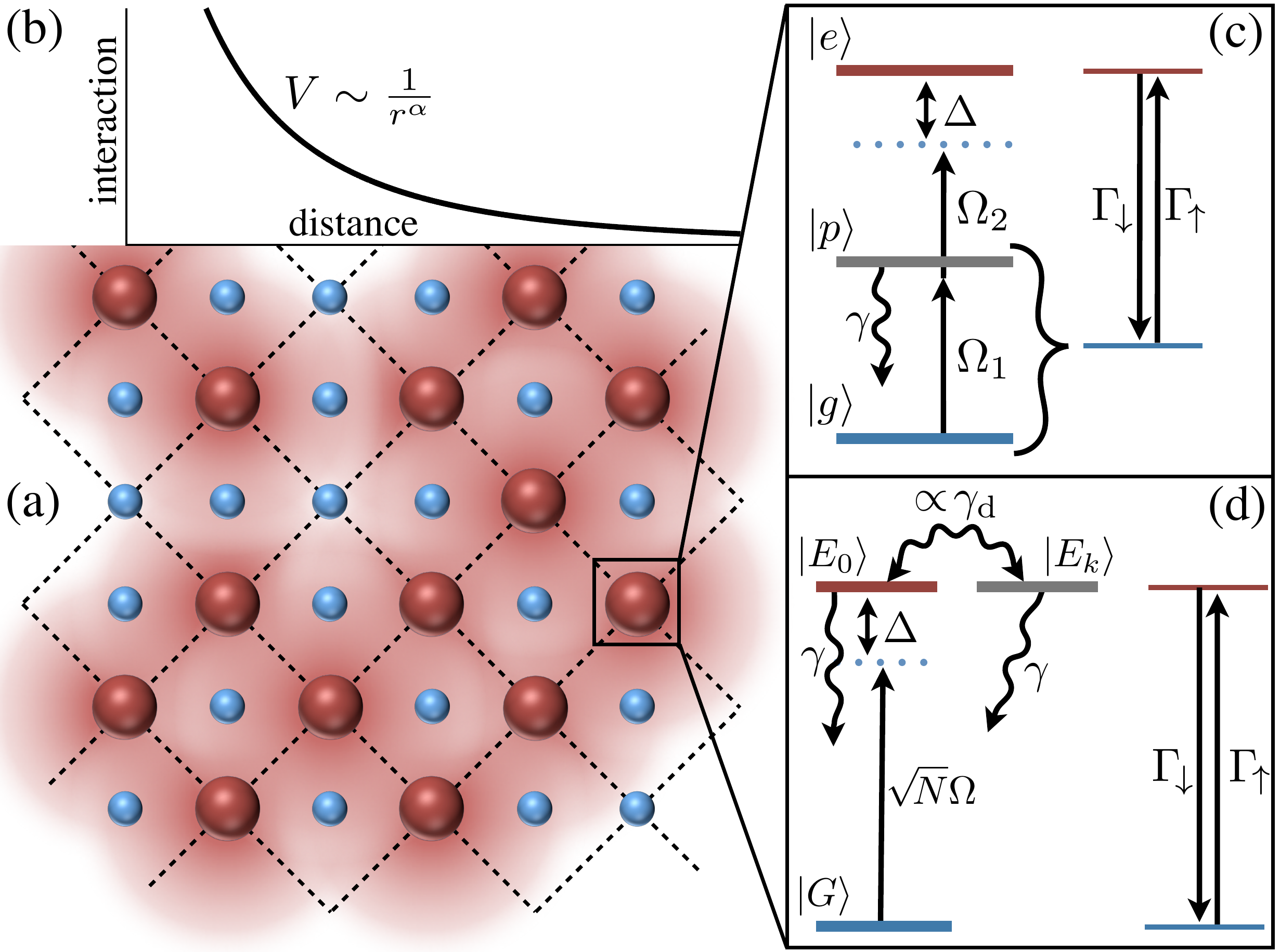}}
\caption{\label{fig1} (color online) (a) Schematics of a two-dimensional lattice in which ground state atoms (small blue spheres) are laser excited to high-lying Rydberg states (large red spheres). The interplay of dissipation and strong Rydberg-Rydberg interactions (b), can give rise to antiferromagnetic long-range order, where excitations predominantly occupy one of the two checkerboard sub-lattices, indicated by the dashed lines in (a). Two possible realizations of such effective two-level systems with tuneable excitation rates, $\Gamma_{\uparrow}$ and $\Gamma_{\downarrow}$, are illustrated in panels (c) and (d) (see text for details).}
\end{center}
\end{figure}
%%%%%%%%%%%%%%%%%%%%%%%%%%%%%%%%%%%%%%%%%%%%%%%%%

In this letter, we address this issue and show that long-range ordered antiferromagnetic phases can indeed be realized in dissipative Rydberg lattices when subject to appropriate coherent driving. However, fluctuations as well as the weak tail of the rapidly decaying interactions are both found to be essential for the physics of the dissipative phase transition. This stands in marked contrast to the equilibrium physics of the corresponding unitary systems, which often is well described by mean field models \cite{wlp08} and NN approximations \cite{jal11}. The crystalline phase features a vanishing dissipative gap and strongly deviates from that of an equilibrium Ising magnet with finite range interactions. Nevertheless, the critical exponent of the non-equilibrium phase transition is found to be that of the Ising universality class.

The unitary evolution is governed by the general Hamiltonian 
\begin{equation}\label{eq:ham}
\hat{H}=\sum_{i}\hat{H}_{i}-\hbar\Delta\sum_{i}\hat{\sigma}_{\rm ee}^{({i})}+V_0\sum_{i<j}\frac{\hat{\sigma}_{\rm ee}^{(i)}\hat{\sigma}_{\rm ee}^{(j)}}{|{\bf r}_i-{\bf r}_j|^{\alpha}},
\end{equation}
for laser-driven atoms on a quadratic lattice of length $L$. The local Hamiltonian $\hat{H}_{i}$ describes the atom-light interaction that excites Rydberg states with a frequency detuning $\Delta$ and $\hat{\sigma}_{\rm ee}^{(i)}$ denotes the corresponding projector onto the Rydberg state of an atom at site ${\bf r}_i=(x_i,y_i)$, $x_i,y_i\in[1,L]$. The last term accounts for the Rydberg-Rydberg interaction, where $V_0=C_{\alpha}/a^{\alpha}$ is the nearest neighbour coupling for a lattice constant $a$ and an interaction strength $C_{\alpha}>0$.  Dipole-dipole interactions correspond to $\alpha=3$ and $\alpha=6$ to van-der-Waals (vdW) interactions. In addition, we consider Markovian loss and dephasing processes described by the operator $\hat{\mathcal{L}}[\rho]$ such that the $N$-body density matrix evolves as $\dot{\rho}=-i[\hat{H},\rho]+\hat{\mathcal{L}}[\rho]$.

For sufficiently strong decoherence, the quantum dynamics of this system can be reduced to the diagonal elements of $\rho$ upon adiabatic elimination of its coherences and neglecting multi-photon excitation of two or more atoms \cite{Ates2007,Ates2011,Heeg2012,hmp13,Les2013,Schonleber2014} \footnote{We have confirmed these simplifications via quantum simulations of smaller lattices.}. This simplifies the time evolution to an effective classical 
rate equation model for the joint probabilities $\rho_{S_1,...,S_N}$ of Rydberg excitations being present ($S_i=1$) or not present ($S_i=0$) at the $i$th site. The corresponding many-body states are connected by single-atom excitation [$\Gamma_{\uparrow}(\delta_i)$] and deexcitation [$\Gamma_{\downarrow}(\delta_i)$] rates
\begin{eqnarray}\label{eq:rate_equ}
&&\dot{\rho}_{S_1,...,S_N}\!\!=\!\!\sum_i \Bigl[(1-S_i)\Gamma_{\downarrow}(\delta_i)+S_i \Gamma_{\uparrow}(\delta_i)\Bigr]\rho_{S_1,...,1-S_i,...,S_N}\nonumber\\
&&\quad -\Bigl[(1-S_i)\Gamma_{\uparrow}(\delta_i)+S_i \Gamma_{\downarrow}(\delta_i)\Bigr]\rho_{S_1,...,S_i,...,S_N}\;,
\end{eqnarray}
where the interactions enter through an effective frequency detuning $\delta_i=\Delta-V_0\sum_{j\neq i}S_j|{\bf r}_i-{\bf r}_j|^{-\alpha}$ that accounts for the level shift of the $i$th atom due to its surrounding Rydberg excitations.

%%%%%%%%%%%%%%%%%%%%%%%%%%%%%%%%%%%%%%%%%%%%%%%%%%%%
\begin{figure}[t!]
\begin{center}
\resizebox{0.99\columnwidth}{!}{\includegraphics{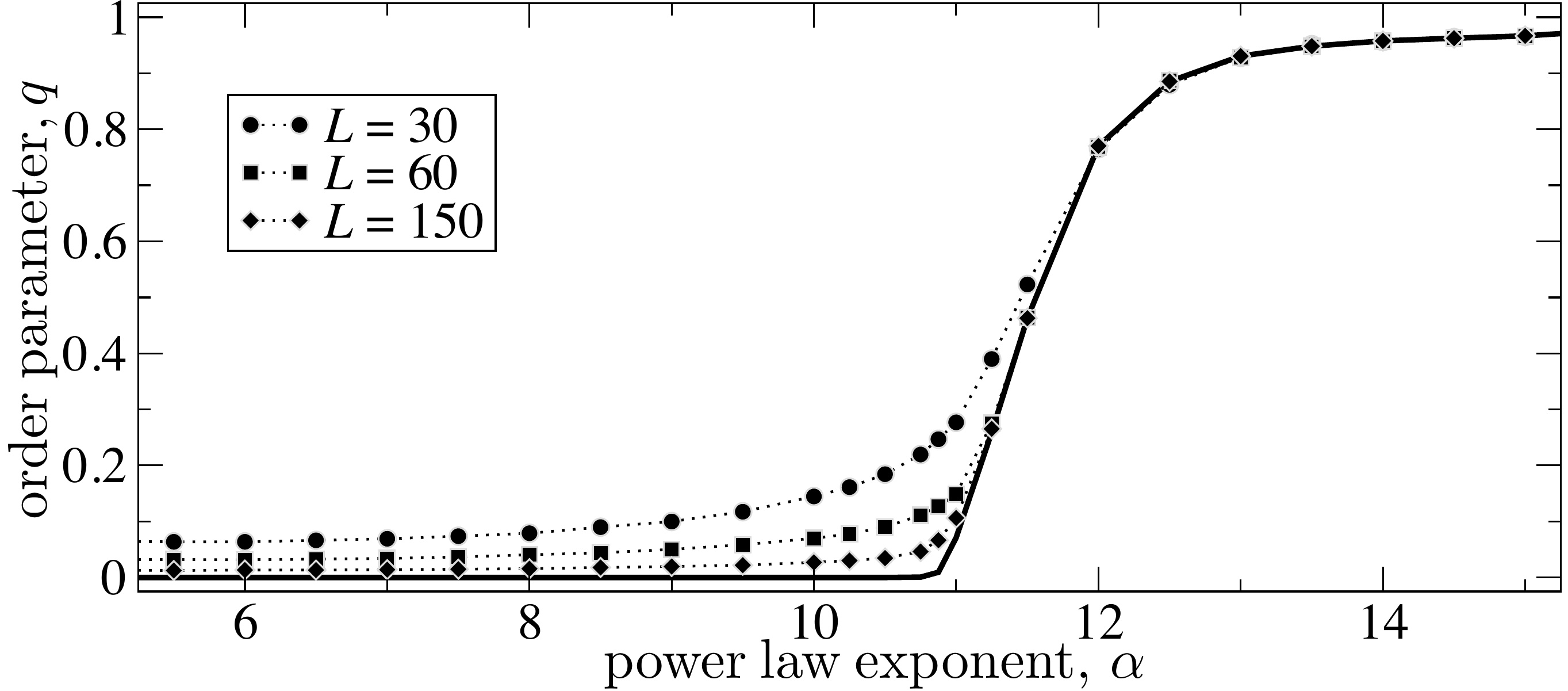}}
\caption{\label{fig2} Order parameter as a function of the power-law exponent $\alpha$, for $p_0=0.95$ and $V_0=5\omega$. Symbols show results for finite system sizes given in the legend. The thick solid line shows the extrapolation to the thermodynamic limit, $L\rightarrow\infty$.}
\end{center}
\end{figure}
%%%%%%%%%%%%%%%%%%%%%%%%%%%%%%%%%%%%%%%%%%%%%%%%%%%%%%%

While the specific form of the rates depends on the particular excitation scheme, the single-atom steady state is given by a simple Lorentzian 
\begin{equation}\label{eq:p0}
\bar{\rho}_{1}(\delta)=\frac{\Gamma_{\uparrow}}{\Gamma_{\uparrow}+\Gamma_{\downarrow}}=\frac{p_0}{1+\delta^2/\omega^2}
\end{equation}
for the settings discussed in here (cf. Fig.\ref{fig1}). Together with the onsite relaxation time $T_1(\delta)$ the rates can be expressed as $\Gamma_\downarrow=(1-\bar{\rho}_{1})/T_1$, $\Gamma_\uparrow=\bar{\rho}_{1}/T_1$. 

We have performed dynamic Monte Carlo (dMC) based on the rates $\Gamma_{\uparrow(\downarrow)}$ and steady-state Monte Carlo (ssMC), assuming $T_1(\delta)=T_1={\rm const.}$, and found good agreement in the relevant parameter regimes. In the latter case, the many-body steady state is  determined by only four parameters: the exponent $\alpha$, the resonant excitation probability $p_0$, the detuning $\Delta/\omega$, and the interaction strength $V_0/\omega$, scaled by the width of the excitation spectrum $\omega$. To detect long-range correlations we define the order parameter
\begin{equation}
q={\Bigl\langle\Bigl|\sum_i(-1)^{x_i+y_i}\hat{\sigma}_{\rm ee}^{(i)}\Bigr|\Bigr\rangle}\Bigl/{\Bigl\langle\sum_i\hat{\sigma}_{\rm ee}^{(i)}\Bigr\rangle}\Bigr.\;.
\end{equation}
As illustrated in Fig.\ref{fig1}, $q$ measures the population imbalance on the two sub lattices reflecting checkerboard ordering, with $q>0$ in the ordered phase and $q=0$ in the disordered phase.

If the Rydberg-Rydberg interaction is approximated by a NN-blockade the above model is analytically solvable, showing that long-range order 
cannot occur in 1D. In higher dimensions, the steady state exhibits N\'eel order provided that $p_0\gtrsim 0.7914$ in 2D and $p_0\gtrsim 0.749$ in 3D square lattices
\cite{Pearce1988}. Thus, for simple two-level driving crystallization is not possible in any dimension since in this case $p_0\leq0.5$. 

In order to clarify the significance of the NN-approximation for crystallization, we have performed ssMC simulations for resonantly driven atoms and varying exponents $\alpha$, for $p_0=0.95$, giving crystallisation under the NN-blockade assumption. As shown in Fig.\ref{fig2}, the NN-approximation fails qualitatively for the important case of vdW interactions ($\alpha=6$), which are found to not support long-range order. Surprisingly, the weak tail of the 
interactions prevents crystallization until a rather large value $\alpha\approx11$. In fact, the simulations show that resonantly driven atoms, with vdW interactions remain in the disordered phase for any values of $p_0$ and $V_0$. 

%%%%%%%%%%%%%%%%%%%%%%%%%%%%%%%%%%%%%%%%%%%%%%%%%%%%%%%
\begin{figure}[t!]
\begin{center}
\resizebox{0.99\columnwidth}{!}{\includegraphics{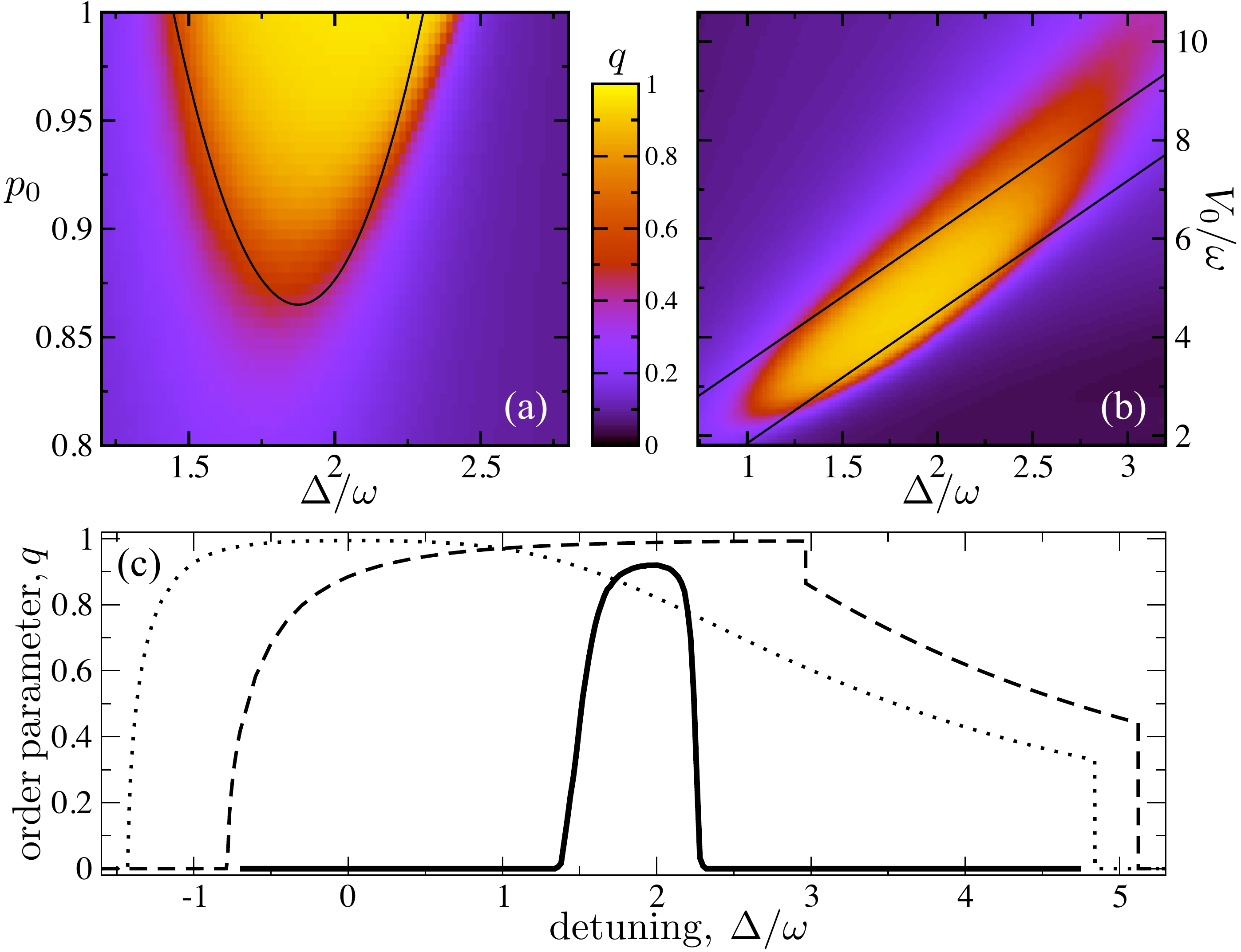}}
\caption{\label{fig3} (color online) Order parameter for a $30\times30$ lattice as a function of (a) $\Delta$ and $p_0$ for $V_0=5$ and (b) $\Delta$ and $V_0$ for $p_0=0.96$. The dashed line shows an estimate of the phase boundary 
(see text). Panel (c) shows the order parameter in the thermodynamic limit for $p_0=0.96$ and $V_0=5$ (solid line) compared to mean field predictions for NN-interactions (dotted line) and the full vdW interactions (dashed line).}
\end{center}
\end{figure}
%%%%%%%%%%%%%%%%%%%%%%%%%%%%%%%%%%%%%%%%%%%%%%%%%%%%%%%%

Long-range order can, however, be stabilized via a finite laser detuning, $\Delta$. This is demonstrated in Fig.\ref{fig3}, showing the order parameter $q$ for finite detunings and a varying $p_0$ and $V_0$. As shown in Fig.\ref{fig3}a, N\'eel-type ordering indeed emerges within a finite detuning range and for $p_0>p_{\rm c}\approx0.86$, only slightly larger than the threshold in the NN-blockade model \cite{Pearce1988}. Yet, N\'eel states are only found in a certain interval of interaction strengths $V_0$, since the vdW tail prevents long-range ordering beyond a critical value (Fig.\ref{fig3}b). 

The location of the transition can be qualitatively understood as follows: A N\'eel state is characterized by a macroscopic population imbalance on the two sublattices with lattice constant $\sqrt{2}a$. Assuming that an atom on the highly populated sub-lattice has an average of $z$ nearest neighbours, the laser detuning must compensate the corresponding level shifts such that its excitation probability remains above threshold, i.e., $\bar{\rho}_1(\Delta/\omega-z V_0/8\omega)\geq p_{\rm c}$, with $\bar{\rho}_1$ given by eq.(\ref{eq:p0}) and $z\approx3$ near the crystallization transition.  The parameter region where this condition is fulfilled is marked in Fig.\ref{fig3} and qualitatively reproduces our numerical results. 

Fig.\ref{fig3}c shows the order parameter as a function of $\Delta$ in the thermodynamic limit, indicating second order phase transitions between the AF and paramagnetic phase. In order to quantitatively assess the importance of fluctuations and the shape of the interaction potential, Fig.\ref{fig3}c also gives a comparison to mean field results under the NN-approximation \cite{lhc11,Qian2012} and for full vdW interactions. Both cases give qualitatively different predictions, suggesting N\'eel order at negative detunings and a first order transition to the reentrant paramagnet at $\Delta>0$.

\begin{figure}[t!]
\begin{center}
\resizebox{0.99\columnwidth}{!}{\includegraphics{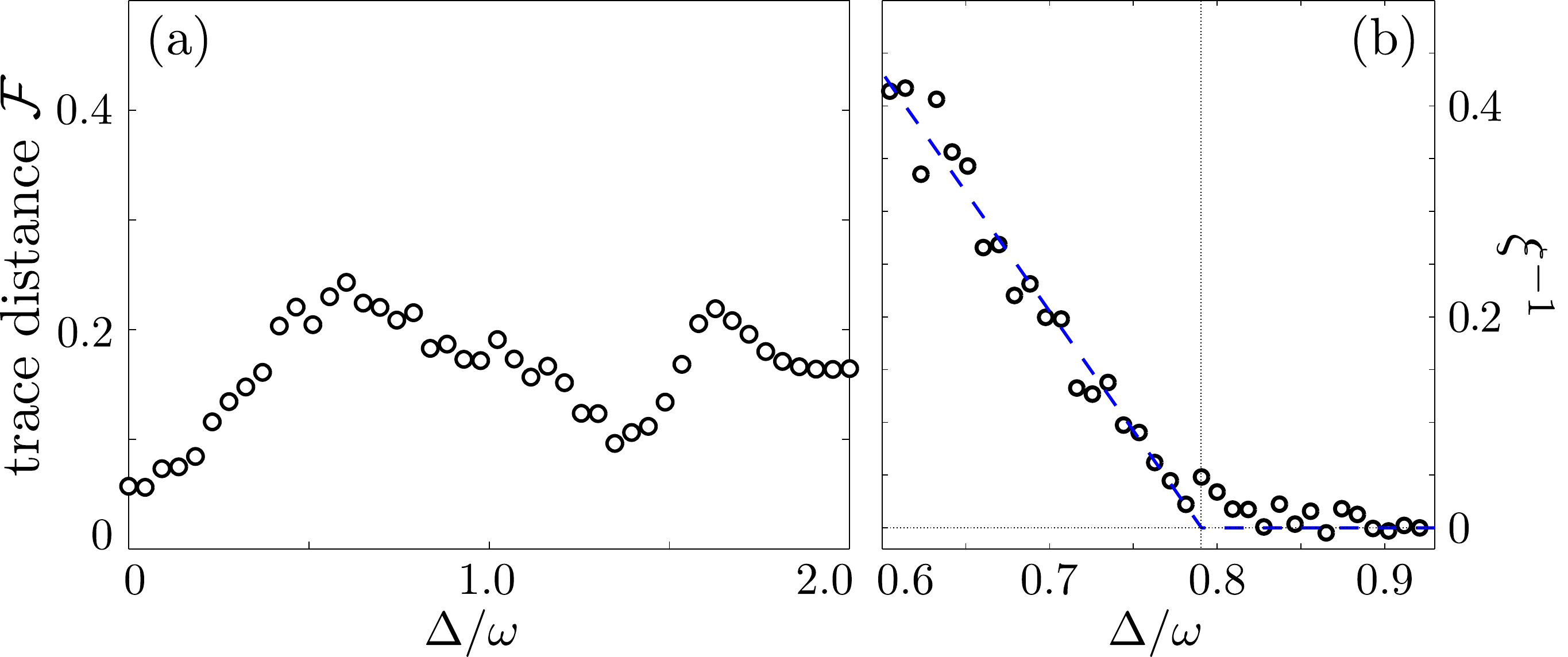}}
\caption{\label{fig4} (a) Trace norm distance ${\cal F}$ of the steady state to a thermal state of an Ising model with optimized 
values of $\beta$, $h$ and $V_{ij}$ [cf. eq.(\ref{eq:rho})] for a $4\times 4$ lattice with periodic boundary conditions and $\alpha=6, p_0=0.9, V_0/\omega=2.97$. 
(b) Finite size extrapolation of the inverse correlation length $\xi^{-1}$ near the 
the phase transition at $\Delta=\Delta_c\approx 0.79 \omega$ for the same set of parameters. The dashed line
corresponds to $\xi^{-1}\propto |\Delta-\Delta_c|^1$.}
\end{center}
\end{figure}

The effects of long-range interactions as well as the dissipative nature of the phase transition can be further illuminated by direct comparison to the corresponding equilibrium situation. In the NN-blockade limit, the steady state, $\bar{\rho}$, of eq.(\ref{eq:rate_equ}) coincides with the thermal equilibrium of a corresponding Ising model 
\begin{equation}
\label{eq:rho}
\rho_{\rm Is}=\frac{1}{Z}\exp\biggl\{-\beta \Bigl[h\sum_j \hat \sigma_j^{z} + \sum_{i<j}V_{i,j} (\hat\sigma_i^{z}+\frac{1}{2})(\hat\sigma_j^{z}+\frac{1}{2})\biggl\}
\end{equation}
where $\hat \sigma^z_i = \hat{\sigma}_{\rm ee}^{(i)}-1/2$, if $V_{i,j}\rightarrow \infty$ for next neighbors and zero otherwise, and $\beta h=\ln \tfrac{1-p_0}{p_0}$. Such a correspondence no longer holds for power-law interactions. This is demonstrated in Fig.\ref{fig4}a, showing the minimum trace-norm distance ${\cal F}=1/2\, {\rm Tr}\{\sqrt{(\bar{\rho}-\rho_{\rm Is})^2}\}$ of the steady state to a thermal state of an Ising model $\rho_{\rm Is}$ with optimized magnetic field, $h$, and interactions, $V_{ij}$.
Note that the average distance between two random density matrices is $0.5$.
Yet, the non-equilibrium steady state exhibits the critical behaviour of the Ising universality class (Fig.\ref{fig4}b), although it differs significantly from that of an Ising model in thermal equilibrium. 

%%%%%%%%%%%%%%%%%%%%%%%%%%%%%%%%%%%%%%%%%%%%%%%%%%%%%%%
%%%%%%%%%%%%%%%%%%%%%%%%%%%%%%%%%%%%%%%%%%%%%%%%%%%%%%%

\begin{figure}[t]
\begin{center}
\includegraphics[width=0.95\columnwidth]{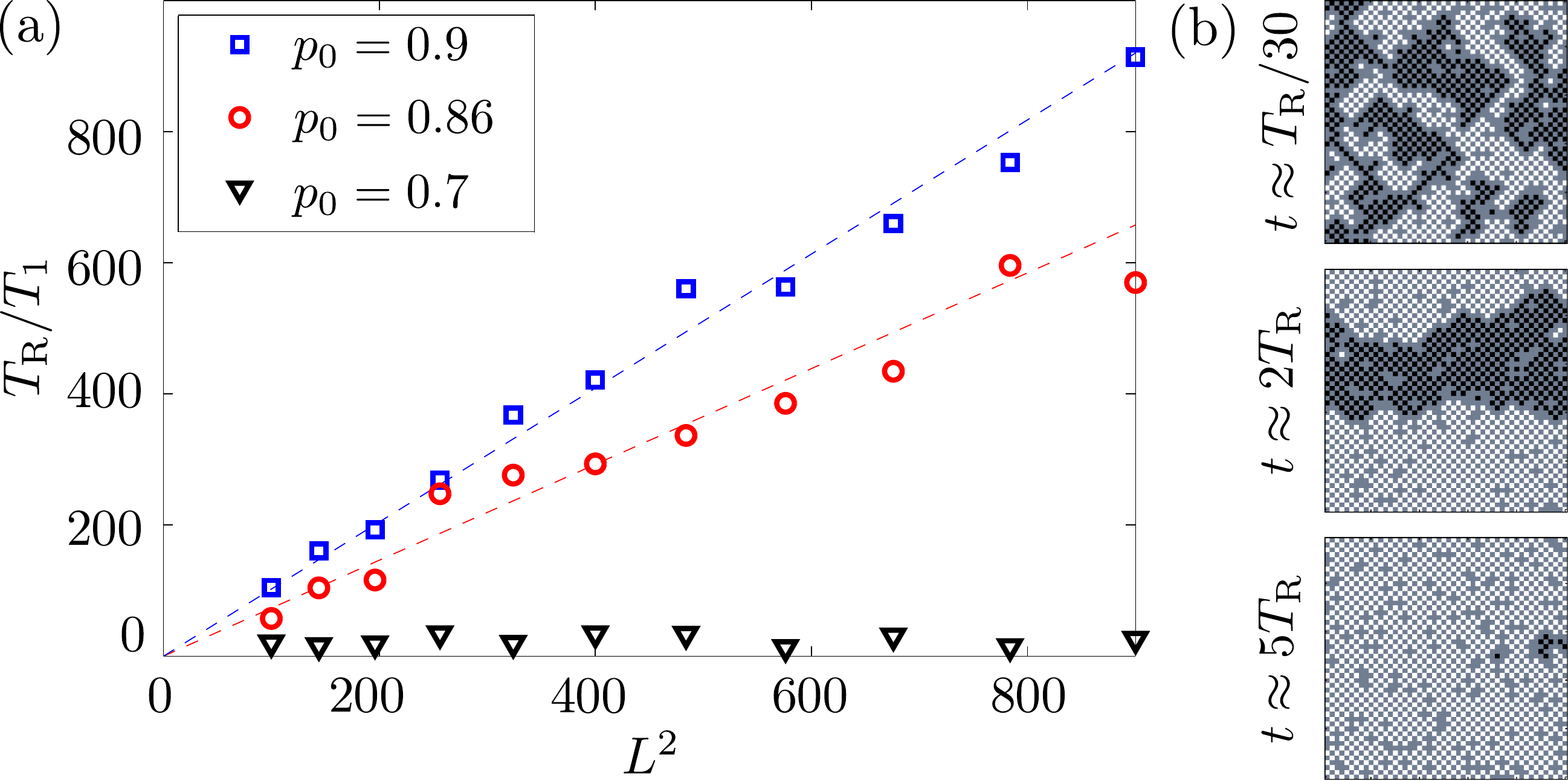}
\caption{(color online). 
(a) Scaling of the relaxation time with system size for a simplified model with NN exclusion. For $p_0=0.7$  
the steady state has no long range order and relaxation does not depend on system size. Contrary, for $p_0=0.86$ and $p_0=0.9$ the AF steady state is reached on a timescale that grows linearly with the number of sites. (b) Snapshots of a single trajectory for $p_0=0.9$ on a $50\times50$ lattice. Sites in the ground state
are marked as grey dots, while excitations on one of the sub-lattices are white and black, respectively.}
\label{fig5} 
\end{center}
\end{figure}

For free dissipative lattice models \cite{Eisert2010} it has been found that long range order is accompanied by a divergence of the system's relaxation time \cite{Honing2012,Horstmann2013}. Using dMC we have access to the full time evolution and the relaxation time. In Fig.\ref{fig5} we show the corresponding results for NN interactions. For $p_0<p_c$, no long range order is established and the relaxation time is independent of system size $L^2$. Crossing the phase transition to the ordered state the relaxation time increases linearly with system size, corresponding to a vanishing dissipative gap in the thermodynamic limit. In Fig.\ref{fig5}b we plot snapshots of a single trajectory for $p_0=0.9$, i.e., well within the ordered regime. Initially, local order emerges quickly, forming small AF domains. The subsequent slow domain wall dynamics leads to a merging into larger clusters, eventually breaking the symmetry and leading to a dominating AF domain on a characteristic timescale $T_{\rm R}$. This final step can be understood as a 1D random walk of the domain walls which gives a relaxation time that scales as $T_\text{R} \sim T_1\, L^2$, reproducing the numerical results of Fig.\ref{fig5}. 

% %%%%%%%%%%%%%%%%%%%%%%%%%%%%%%%%%%%%%%%%%%%%%%%%%%%%%%%%%
% %%%%%%%%%%%%%%%%%%%%%%%%%%%%%%%%%%%%%%%%%%%%%%%%%%%%%%%%%

We finally discuss two experimental implementations of the described system that overcome the two-level limit of $p_0\leq0.5$. Consider first a lattice of small atomic clouds, each containing $N$ two-level atoms. Here $\hat{H}_i=\Omega\hat\sigma_\mathrm{eg}^{(i)}+h.c.$ couples directly the ground and Rydberg state, which decays with a spontaneous decay rate $\gamma$. For sufficiently strong Rydberg interactions within the cloud, the corresponding interaction blockade inhibits multiple Rydberg excitations, such that each site acts like an effective two-level "super-atom" \cite{Lukin2001} composed of the collective ground state $|G\rangle$ and the symmetric, singly excited state $|E_0\rangle$ with an enhanced Rabi frequency $\sqrt{N}\Omega$. Additional single-atom dephasing with a rate $\gamma_{\rm d}$ transfers population between $|E_0\rangle$ and the manifold of $N-1$ singly excited, non symmetric states $|E_k\rangle$ \cite{Honer2011} (Fig. \ref{fig1}d). In the limit of strong dephasing, adiabatic elimination yields classical rates $\Gamma_{\uparrow(\downarrow)}$ for transitions between $|G\rangle$ and the manifold of singly excited states. The steady state excitation probability is given by eq.(\ref{eq:p0}) with $p_0=N / [N(1+\gamma/\bar{\gamma})+\gamma_d/\bar{\gamma}+\gamma\bar{\gamma}/(4\Omega^2)]$, where $\bar{\gamma}=\gamma+\gamma_{\rm d}$. For large $N$, $p_0\rightarrow1/(1+\gamma/\gamma_{\rm d}+1/N)$, which approaches unity for strong dephasing \cite{Honer2011} and $\omega=\sqrt{N}\Omega\sqrt{\gamma_\mathrm{d}/\gamma}$. Recent experiments have studied dissipation effects in random frozen Rydberg gases \cite{Malossi2013}. The described super-atom lattices can be realized with magnetic or optical micro trap arrays \cite{Weinstein1995,Dumke2002,Whitlock2009} that  accommodate $N\approx10\dots10^2$ atoms per site and provide lattice constants of a few $\mu$m for which strong NN interactions can be obtained.

Alternatively $p_0$ can be controlled on a single-atom level using three-level excitation of Rydberg atoms via a low-lying intermediate state, $|p\rangle$, with two Rabi frequencies $\Omega_1$ and $\Omega_2$ (Fig.\ref{fig1}c). Here, the fast decay of the intermediate state with a rate $\gamma\sim $MHz drives the relaxation towards the steady state eq.(\ref{eq:p0}), with a tuneable $p_0=\Omega_1^2/(\Omega_1^2+\Omega_2^2)$. Such three-level excitation schemes are utilised in numerous Rydberg atom experiments, either exploring interaction effects in the strong excitation regime ($\Omega_1>\Omega_2$) \cite{Schempp2010,Schwarzkopf2013,Nipper2012,Viteau2012} or quantum optics applications in the opposite limit \cite{Pritchard2010,Dudin2012,Peyronel2012,Maxwell2013}. As a specific example, laser excitation of Rb($35S_{1/2}$) Rydberg states via the intermediate Rb($5P_{1/2}$) state with $\Omega_1=0.5\gamma=4\Omega_2$ yields $p_0\approx0.9$, upon accounting for the small but non-negligible Rydberg state decay. For a lattice constant of $a\approx2\mu$m these conditions correspond to $V_0\approx5\omega$, i.e. well within the parameter region of the ordered steady state. Rydberg excitation and trapping \cite{Anderson2011} as well as site-resolved Rydberg atom imaging \cite{Schauss2012} has been experimentally demonstrated in 2D lattices with $a\approx0.5\mu$m. Larger lattice constants can also be realized in these settings \cite{Fukuhara2013} or via single-atom trapping in optical micro-trap arrays \cite{Beguin2013}, such that the creation and probing of the predicted dissipative phase transition appears to be well within experimental reach.

In conclusion, we have shown that Rydberg lattices can undergo a dissipative phase transition to a long range ordered AF phase. A key requirement is the use of optical coupling schemes that go beyond the inversion limit of simple two level driving and a finite laser detuning to counteract the effects of the power-law tail of the interaction potential. While we have focussed here on neutral-atom settings with \emph{finite-range} vdW interactions ($\alpha=6$), effective quantum magnets with variable power-law interactions \cite{Schach2013} are currently attracting great interest in the context of laser-cooled ion crystals in which various spin models \cite{Porras2004} with $\alpha=0\ldots3$ can be realized in one and two dimensions \cite{Britton2012,Islam2013,Richerme2014}. These systems inherently feature dissipation \cite{Foss2013} and, thus, provide an interesting platform to explore the revealed dissipative phase transition also in the \emph{long-range} interaction regime. In light of the demonstrated failure of mean field approximations, it would further be of great interest to investigate other dissipative phase transitions predicted by mean field theory \cite{Lee2013} and gain insight into its validity for open systems as well as the critical dimension for long-range order in such related spin lattices. The present phase transition was found to fall into the Ising universality class, despite marked differences of its steady states from an Ising model in thermal equilibrium, raising the general question of how a critical theory of phase transitions \cite{Eisert2010} in interacting dissipative settings relates to that of thermal or quantum phase transitions. We have addressed here the limit of strong decoherence, and understanding the transition behaviour to weakly damped, strongly interacting quantum many-body systems poses a challenging and likewise important issue for future work.

\begin{acknowledgments} 
We thank C. O'Brien, D. Petrosyan, J. Otterbach, M. G\"arttner and J. Evers for discussions and Georg Bannasch for valuable contributions in the early stages of this work. Financial support by the Deutsche Forschungsgemeinschaft through SFB TR49 and by the EU through the Marie Curie ITN "COHERENCE" and EU-FET Grant No. HAIRS 612862 is acknowledged. 
\end{acknowledgments}


\begin{thebibliography}{99}
\bibitem{bdz08} I. Bloch, J. Dalibard and W. Zwerger, Rev. Mod. Phys. {\bf 80}, 885 (2008).
\bibitem{res_meas} T. Gericke \emph{et\ al.} Nature Phys. {\bf 4}, 949 (2008); N. Gemelke \emph{et\ al.}, Nature {\bf 460}, 995 (2009); J. F. Sherson \emph{et\ al.}, Nature {\bf 467}, 68 (2010); W. S. Bakr, \emph{et\ al.}, Science {\bf 329}, 547 (2010).
\bibitem{ryd_rev} M. Saffman, T. G. Walker and K. Molmer, Rev. Mod. Phys. {\bf 82}, 2313 (2010).
\bibitem{Lahaye2009} T. Lahaye, C. Menotti, L. Santos, M. Lewenstein, T. Pfau, Rep. Prog. Phys. {\bf 72}, 126401 (2009)
\bibitem{pdl10} T. Pohl, E. Demler, and M. D. Lukin, Phys. Rev. Lett. {\bf 104}, 043002 (2010).
\bibitem{slm10} J. Schachenmayer, I. Lesanovsky, A. Micheli and A. J. Daley, New J. Phys. {\bf 12}, 103044 (2010).
\bibitem{bsl11} R. M. W. van Bijnen \emph{et\ al.}, J. Phys. B {\bf 44}, 184008 (2011).
\bibitem{lhc11} T. E. Lee, H. H\"affner and M. C. Cross, Phys. Rev. A {\bf 84}, 031402 (2011).
\bibitem{Diehl2010} S. Diehl \emph{et\ al.}, Phys. Rev. Lett. {\bf 105}, 015702 (2010).
\bibitem{Mueller2012} M. M\"uller \emph{et\ al.} Adv. Atom. Mol. Opt. Phys. {\bf 61}, 1  (2012).
\bibitem{Sieberer2013} L. Sieberer \emph{et\ al.}, Phys. Rev. Lett. {\bf 110}, 195301 (2013).
\bibitem{Lee2013} T.E. Lee, S. Gopalakrishnan, and M.D. Lukin, Phys. Rev. Lett. {\bf 110}, 257204 (2013).
\bibitem{Ates2012} C. Ates, B. Olmos, J. P. Garrahan, and I Lesanovsky, Phys. Rev. A {\bf 85}, 043620 (2012).
\bibitem{hmp13} M. H\"oning, D. Muth, D. Petrosyan and M. Fleischhauer, Phys. Rev. A {\bf 87}, 023401 (2013).
\bibitem{Qian2012} J. Qian, G. Dong, L. Zhou, and W. Zhang, Phys. Rev. A {\bf 85}, 065401 (2012).
\bibitem{wlp08} H. Weimer, R. L\"ow, T. Pfau and H. P. B\"uchler, Phys. Rev. Lett. {\bf 101}, 250601 (2008).
\bibitem{jal11} S. Ji, C. Ates, and I. Lesanovsky, Phys. Rev. Lett. {\bf 107}, 060406 (2011).
\bibitem{Ates2007} C. Ates, T. Pohl, T. Pattard, and J. M. Rost, Phys. Rev. Lett. {\bf 98}, 023002 (2007); Phys. Rev. A {\bf 76}, 013413 (2007); J. Phys. B {\bf 39}, L233 (2006).
\bibitem{Ates2011} C. Ates, S. Sevincli, and T. Pohl, Phys. Rev. A {\bf 83}, 041802 (2011).
\bibitem{Heeg2012} K. P. Heeg, M. G\"arttner, and J. Evers, Phys. Rev. A {\bf 86}, 063421 (2012).
\bibitem{Les2013} I. Lesanovsky, and J. P. Garrahan, Phys. Rev. Lett. {\bf 111}, 215305 (2013).
\bibitem{Schonleber2014} D. W. Sch\"onleber, M. G\"arttner, J. Evers, arXiv:1401.7260.
\bibitem{Pearce1988} P. Pearce and K. Seaton, J. Stat. Phys. {\bf 53}, 1061 (1988).
\bibitem{Eisert2010} J. Eisert and T. Prosen, arXiv:1012.5013.
\bibitem{Honing2012} M. H\"oning, M. Moos, and M. Fleischhauer, Phys. Rev. A {\bf 86}, 013606 (2012).
\bibitem{Horstmann2013} B. Horstmann, J. I. Cirac and G. Giedke, Phys. Rev. A {\bf 87}, 012108 (2013).
\bibitem{Lukin2001} M.D. Lukin \emph{et\ al.}, Phys. Rev. Lett. {\bf 87} 037901 (2001).
\bibitem{Honer2011} J. Honer, R. L\"ow, H. Weimer, T. Pfau, and H.P. B\"uchler, Phys. Rev. Lett. {\bf 107}, 093601 (2011).
\bibitem{Malossi2013} N. Malossi \emph{et\ al.}, arXiv:1308.1854.
\bibitem{Weinstein1995} J. D. Weinstein and K. G. Libbrecht, Phys. Rev. A {\bf 52}, 4004  (1995). 
\bibitem{Dumke2002} R. Dumke \emph{et\ al.}, Phys. Rev. Lett. {\bf 89}, 097903 (2002) 
\bibitem{Whitlock2009} S. Whitlock, R. Gerritsma, T. Fernholz and R. J. C. Spreeuw, New J. Phys. {\bf 11} 023021 (2009).
\bibitem{Schempp2010} H. Schempp, \emph{et\ al.}, Phys. Rev. Lett. {\bf 104}, 173602 (2010); Phys. Rev. Lett. {\bf 112}, 013002 (2014).
\bibitem{Schwarzkopf2013} A. Schwarzkopf \emph{et al.}, Phys. Rev. Lett. {\bf 107}, 103001 (2011); Phys. Rev. A {\bf 88}, 061406 (2013).
\bibitem{Nipper2012} J. Nipper \emph{et al.}, Phys. Rev. X {\bf 2}, 031011 (2012).
\bibitem{Viteau2012} M. Viteau \emph{et al.}, Phys. Rev. Lett. {\bf 109}, 053002 (2012).
\bibitem{Pritchard2010} J. D. Pritchard \emph{et al.}, Phys. Rev. Lett. {\bf 105} 193603 (2010).
\bibitem{Dudin2012} Y.O. Dudin A. Kuzmich, Science {\bf 336}, 887 (2012).
\bibitem{Peyronel2012} T. Peyronel \emph{et\ al.}, Nature {\bf 488}, 57 (2012).
\bibitem{Maxwell2013} D. Maxwell \emph{et\ al.}, Phys. Rev. Lett. {\bf 110}, 103001 (2013).
\bibitem{Anderson2011} S. E. Anderson, K. C. Younge, and G. Raithel, Phys. Rev. Lett. {\bf 107}, 263001 (2011).
\bibitem{Schauss2012} P. Schauss \emph{et\ al.}, Nature {\bf 491}, 87 (2012).
\bibitem{Fukuhara2013} T. Fukuhara \emph{et\ al.}, Nature {\bf 502}, 76 (2013).
\bibitem{Beguin2013} L. B\'eguin \emph{et\ al.}, Phys. Rev. Lett. {\bf 110}, 263201 (2013); D. Barredo \emph{et\ al.}, arXiv:1402.4077; F. Nogrette \emph{et\ al.}, arXiv:1402.5329.
\bibitem{Schach2013} J. Schachenmayer, B. P. Lanyon, C. F. Roos and A. J. Daley, Phys. Rev. X {\bf 3}, 031015 (2013).
\bibitem{Porras2004} D. Porras and J.I. Cirac, Phys. Rev. Lett. {\bf 92}, 207901 (2004).
\bibitem{Britton2012} J. W. Britton \emph{et\ al.}, Nature 484, 489-492 (2012).
\bibitem{Islam2013} R. Islam \emph{et\ al.}, Science {\bf 340}, 583 (2013).
\bibitem{Richerme2014} P. Richerme \emph{et\ al.}, arXiv:1401.5088; P. Jurcevic \emph{et\ al.}, arXiv:1401.5387.
\bibitem{Foss2013} M. Foss-Feig \emph{et\ al.}, Phys. Rev. A {\bf 87}, 042101 (2013);New J. Phys. {\bf 15}, 113008 (2013).
\end{thebibliography}
\end{document}